\documentclass[a4paper,10pt,twoside]{cpc-hepnp}

\usepackage{multicol}
\usepackage{graphicx}
\usepackage{booktabs}
\usepackage{amssymb,bm,mathrsfs,bbm,amscd}
\usepackage[tbtags]{amsmath}
\usepackage{lastpage}

\def\ie{{\it i.e. }}

\def\etal{{\it et al. }}

\def\PRD{{\em Phys. Rev.} D}

\def\NPA{{\em Nucl. Phys.} A}
\def\NPB{{\em Nucl. Phys.} B}
\def\PHR{{\em Phys. Rev.} }
\def\PLB{{\em Phys. Lett.} B}
\def\JHP{{\em JHEP} }

\def\PRC{{\em Phys. Rev.} C}
\def\PRL{{\em Phys. Rev. Lett.} }

\begin{document}

\fancyhead[c]{\small Chinese Physics C~~~Vol. xx, No. x (201x) xxxxxx}

\footnotetext[0]{Received 20 August 2013}

\title{The meson polarized distribution
function and  mass dependence of the nucleon parton densities }

\author{
\quad A.Mirjalili$^{a;1)}$\email{a.mirjalili@yazd.ac.ir}\;\;\;    and \quad K. Keshavarzian$^{a)}$
} \maketitle

\address{%
$^{a}$Physics Department, Yazd University, 89195-741,Yazd, Iran }

\begin{abstract}
The polarized distribution functions of mesons, including pion,
kaon and eta, using the proton structure function, are calculated.
We are looking for a relationship between the polarized distribution
of mesons and the polarized structure of nucleons. We show that the
meson polarized parton distributions leads to zero total spin
for mesons, considering the orbital angular momentum of quarks and
gluons inside the meson. Two separate Monte Carlo algorithms are
applied to compute the polarized parton distributions of the kaon.
Via the   mass dependence of quark distributions, the distribution
function of the eta  meson is obtained. A new method by which the
polarized sea quark distributions of protons are evolved separately
-- which cannot be performed easily using the standard solution
of DGLAP equations -- is introduced. The mass dependence of these
distributions is obtained, using the renormalization group
equation which makes their evolutions more precise.
Comparison between the evolved distributions and the available
experimental data validates the suggested solutions for separate
evolutions.
\end{abstract}



\footnotetext[0]{\hspace*{-3mm}\raisebox{0.3ex}{$\scriptstyle\copyright$}2013
Chinese Physical Society and the Institute of High Energy Physics
of the Chinese Academy of Sciences and the Institute
of Modern Physics of the Chinese Academy of Sciences and IOP Publishing Ltd}%

\section{Introduction}The effective Lagrangian has been written by Manohar
\cite{manohar} to justify the chiral symmetry of quarks. This
model can be extended to polarized parton distributions and
many studies have used it to find these distributions for quarks
inside nucleons \cite{weise}. There are also many studies which
calculate the polarized parton distribution function (PPDF) of mesons,
based on lattice QCD computations \cite{lattice} or other
approaches \cite{ma,namkim}. We know that due to the
orbital angular momentum of quarks and gluons inside hadrons, the
$+$ and $-$ helicity distributions do not compensate each other
exactly \cite{ji}. If we extend our theoretical  framework to the case where the
meson mass corrections and higher twist effects are considered,
then it is possible to consider the longitudinal polarization for
pion parton distribution. It turns out that the transverse
polarization, which is denoted  by $g_2$ \cite{dong}, is connected
with the longitudinal polarization structure $g_1$ via the
Wandzura-Wilczek relation \cite{ww}. In a similar fashion, we can
also consider some extra effects due to meson-mass correction
which will lead us to additional longitudinal polarization for
the partons of pseudo-scalar mesons. The structure of meson-mass
corrections in inclusive processes is in general more complicated
than that of target-mass corrections in deep inelastic scattering,
which can be re-summed using the Nachtmann variable. The twist
approximation which is used for the amplitude distribution of pions
can also be employed in deep inelastic
scattering processes \cite{ball}. In addition to this theoretical
justification for assigning longitudinal polarization to the partons of
pseudo-scalar mesons, we
can also consider the diffraction effect, using the factorization
theorem for the hard exclusive electro-production of mesons in QCD.
The full theorem applies to all kinds of meson and not just to
vector mesons. The parton densities used include not only the
ordinary parton densities, but also the helicity densities.\\

In this article we  try to calculate the
PPDF of mesons using the (definite) PPDF of nucleons. This work contains two separate parts:\\

1) Computing the polarization densities and orbital angular momentum
of quarks and gluon inside the meson.

2) Evolving the sea quark distributions of nucleons (in which their
symmetry is broken) separately, using  the renormalization group
equation
for the running mass of quarks.\\

In Part 1, we  calculate the bare quark distributions using the proton
polarized structure function  $g_1^p(x)$, using data from~\cite{e143}. Then
we  compute the ratio of the polarized valence data of kaons to that
of pions, $\delta q_{val}^K / \delta q_{val}^{\pi}$, using the data
for their  unpolarized ratio, $q_{val}^K /
q_{val}^{\pi}$ \cite{badier}, based on two separate Monte Carlo
algorithms. We also calculate the polarized valence ratio of eta
mesons to pions, $\delta q_{val}^{\eta} / \delta q_{val}^{\pi}$,
using the mass dependence of the valence quark distribution inside the
meson. Substituting these ratios into the chiral quark model
($\chi$QM) equations and fitting with experimental data (or any
reasonable phenomenological model), the polarized distribution
functions in pion, kaon and eta mesons at low energy scales will
be obtained. Following that, the evolution of the PDFs, employing the DGLAP
equations, can be done  straightforwardly  \cite{hwa1,hwa2,hwa3,hwa4}. Using
these evolved PDFs, we can extract the values of the orbital angular
momentum of quarks and gluons inside mesons \cite{ji}.

In Part 2, the PPDFs of the proton, using the distributions extracted
for mesons, are calculated. The valence PPDF of the proton can be
evolved easily using the non-singlet moment $\delta M_{NS}$.
Since the DGLAP equations can thoroughly evolve only the sea quark
distribution, however, the evolution of the separated sea quarks
is more complicated. There are reasonable methods to separate the
evolution of sea quarks \cite{GRSV} but in this work we use the
running mass and renormalization equation to make the sea quark
distribution functions depend on the quark masses. Thereby, the
eigenvalues of the evolution operator become non-degenerate. The
sea quark distributions at low scale $Q^2_0$, arising from $\chi
QM$, are unsymmetrized. The different eigenvalues of the evolution
operator, which are obtained as a result of the new
method introduced in this paper, makes the evolution of sea quark densities
more distinctive than what we obtained in ~\cite{MKY-2011}. Two
boundary conditions at low- and high-energy scales are applied to
the equations to test the sea quark spectrum. Finally, a
comparison to experimental data is carried out for sea and valence
distributions \cite{hermes,smc}.

This paper is organised as follows. In Section 2 we review the
basic concepts of $\chi$QM in the unpolarized case. The extension of
this model to the polarized case is done in Section 3. In Section
4 we deal with a method to extract the polarized bare quark
distributions inside the proton. In Section 4.1, two Monte Carlo
algorithms are introduced which give us the polarized valence
distributions of the kaon and in Section 4.2 we calculate the
distribution function of the eta meson, using the fact that
the masses of the quarks gives different  distributions for the various quark flavours. The parton
orbital angular momentum inside the meson and the spin of meson is
discussed in Section 5. In Section 6 we use the renormalization
group equation for the running mass of quarks to get the separated
evolution operators for nucleon sea quark densities. We give our
conclusions in Section 7.

\section{Unpolarized chiral quark model}
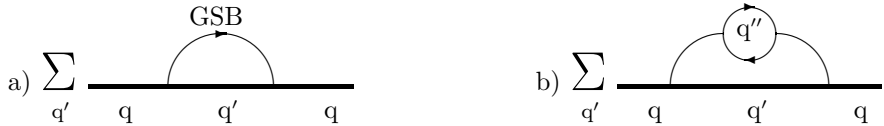
\begin{figure}[htp]
\begin{center}
\begin{picture}(400,80)(0,0)
\linethickness{0.4 mm} \put(50,33){\line(1,0){100}}
\put(250,33){\line(1,0){100}} \linethickness{0.2 mm}
\put(15,25){\makebox(35,25){a) \huge{$\Sigma$}}}
\put(215,25){\makebox(35,25){b) \huge{$\Sigma$}}}
\put(100,33){\oval(40,40)[t]} \put(290,33){\oval(40,40)[tl]}
\put(310,33){\oval(40,40)[tr]} \put(300,53){\circle{20}}
\put(100,53){\vector(1,0){2}} \put(300,63){\vector(1,0){2}}
\put(300,43){\vector(-1,0){2}}
\put(36,17){\makebox(315,13)[l]{{\footnotesize q$'$} \hspace*{4
mm} q \hspace*{9 mm} q$'$ \hspace*{9 mm} q \hspace*{30 mm}
{\footnotesize q$'$} \hspace*{4 mm} q \hspace*{9 mm} q$'$
\hspace*{9 mm} q}} \put(88,53){\makebox(25,13)[l]{GSB}}
\put(295,48){\makebox(13,13)[l]{q$''$}}
\end{picture}

\caption{At low energy, the bare quarks are dressed as
indicated by diagram (a). At higher energies, probing reveals the
structure of GS bosons (pion, kaon and eta)  as is shown in
diagram (b). This figure has been adapted from ~\cite{MKY-2011}.
\label{pros}}
\end{center}
\end{figure}

Our calculations are based on the constituent quark Fock state
using the chiral quark model, $\chi$QM \cite{manohar}. According
to this model, spontaneous  chiral symmetry breaking creates
Goldstone (GS) bosons  which couple to the constituent quarks. The
low-energy dynamics ($\mu \leq 4\pi f_{\pi}\sim 1$ GeV where
$f_{\pi}\simeq 93$ MeV is the pion decay constant) is governed by
the GS bosons, in particular the pion, which is the approximate
zero mode of QCD vacuum \cite{manohar,weise,urek}. The diagrams
responsible for this process are as shown in
Fig.~{\ref{pros}}.

The interaction Lagrangian of the effective chiral quark theory in
the leading order of an expansion in $\Pi/f$ is given by \cite{manohar}:
\begin{equation}
\mathcal{L}=-{g_A \over
f}\bar{\psi}(\partial_{\mu}\Pi)\gamma^{\mu}\gamma_5\psi,\label{lagrange}
\end{equation}
while the GS boson matrix field is written  as:
\begin{equation}
\Pi={1 \over \sqrt{2}} \left(
\begin{array}{ccc}
{\pi^0 \over \sqrt{2}}+{\eta \over \sqrt{6}} & \pi^+ & K^+ \\
\pi^- & -{\pi^0 \over \sqrt{2}}+{\eta \over \sqrt{6}} & K^0 \\
K^- & \bar{K}^0 & -{2\eta \over \sqrt{6}} \\
\end{array}
\right). \label{matrix}
\end{equation}
Using the notation of Ref.~\cite{weise}, we can write the
constituent U and D quark Fock-state as:
\begin{equation}
|q\rangle=\sqrt{Z}|q_0\rangle + \sum_M
\alpha_{M}|qM\rangle,\label{constit}
\end{equation}
where Z is the renormalization constant for a ``bare'',
$|q_0\rangle$, constituent quark and we have absorbed all
coefficients in $\alpha_{M}$. The $|qM\rangle$ indicates the quark
states dressed by GS bosons. Eq.~{(\ref{constit})} will yield the
one-point Fock state contribution and ignores higher order
approximations (see Fig.~{\ref{pros}}).

In the unpolarized case, the splitting function which gives the
probability to convert a parent constituent quark $q$ into a
constituent quark $q'$ carrying the light-cone momentum fraction
$(1-x_M)$, and a spectator GS boson (pion, koan, eta) carrying the
momentum fraction $x_M$, is given by \cite{weise}:
\begin{eqnarray}
f_{q/q'M} & = & \left( {g_{q/q'M} \over 4\pi}\right)^2 {1\over
x_M(1-x_M)^2}\int_0^\infty dk_\bot^2\left| G_{q/q'M}\right|^2
{[(1-x_M)m_q-m_{q'}]^2+k_\bot^2 \over (m_q^2-M_{q'M}^2)^2}, \nonumber \\
&& \label{spli}
\end{eqnarray}
where
\begin{equation}
g_{q/q'M}={g_A\over f}\bar{m},\qquad \bar{m}={m_q+m_{q'}\over
2},\label{g}
\end{equation}
and the vertex function can be written as:
\begin{equation}
G_{q/q'M}=\exp \left( {m_q^2-M_{q'M}^2\over
2\Lambda^2}\right).\label{gg}
\end{equation}
In Eq.~{(\ref{g})}, $f$ is the pseudo-scalar decay constant and is
taken to be equal to the pion decay constant, so $f\approx 93$ MeV. The
quark axial-vector coupling is represented by $g_A$ and it can
be taken to be $1$, as suggested in Ref.~\cite{weinberg}, or $0.75$,
as suggested in Ref.~\cite{manohar}. In our calculations below we
choose the former value.  The cut-off parameter is  $\Lambda$ and
 is usually determined
phenomenologically \cite{manohar,weise,urek,kshvzn}. We use its
previously determined value, $\Lambda=1.4\ \mathrm{GeV}$
\cite{weise}.

In Eq.~{(\ref{gg})}, $G_{q/q'M}$ is the vertex function and accounts
for the extended structure of the GS bosons and the constituent
quark. $M_{q'M}^2$ is the invariant mass squared of the
\textquotedblleft meson + constituent quark \textquotedblright\
system \cite{weise,urek}:
\begin{equation}
M_{q'M}^2={M_M^2+k_\bot^2 \over x_M}+{m_{q'}^2+k_\bot^2 \over
1-x_M}.\label{mm}
\end{equation}
In Eq.~(\ref{mm}), $m_{q'}$ denotes the mass of constituent quark $q'$. In our
calculations we  use $m_u=m_d=360\ \mathrm{MeV}$ and $m_s=570\
\mathrm{MeV}$ as typical values guided by NJL model
calculations \cite{nambu}.

\section{Chiral quark model in the  polarized case}
\begin{figure}[htp]
\begin{center}
\includegraphics[width=8.5 cm]{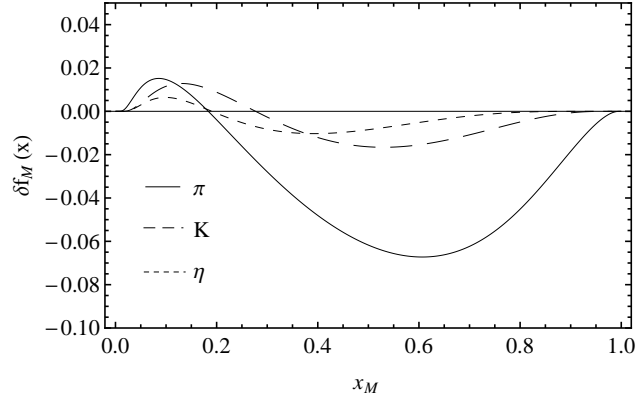}
{\caption{The polarized  splitting functions  $\delta f_{q/q'M}$
for  pion, kaon and eta. \label{figspli}}}
\end{center}
\end{figure}

Although the parton picture only applies to high energy processes, it is possible to obtain  the parton densities at low energy scales using the effective Lagrangian. In the following,  we need  to employ the chiral quark model in which the bare quarks are surrounded by meson  clouds. The result of this approach is that we can access the constituent U and D quarks, which lead us to achieve the parton densities of the nucleon at low energy scales. To calculate the nucleon PPDFs in the chiral quark model, the polarized splitting function is needed \cite{manohar,weise,urek,kshvzn}:
\begin{eqnarray}
\delta f_{q/q'M} & = & \left( {g_{q/q'M}\over
4\pi}\right)^2{1\over x_M(1-x_M)^2}\int_0^\infty dk_\bot^2\left|
G_{q/q'M}\right|^2{[(1-x_M)m_q-m_{q'}]^2 -k_\bot^2\over
(m_q^2-M_{q'M}^2)^2}, \nonumber \\
&& \label{polspli}
\end{eqnarray}
which is analogous to Eq.~{(\ref{spli})} with the exception of the
minus sign before $k_\bot^2$. The expression $f_{q/q'M}=f_{q/q'M}\uparrow
+f_{q/q'M}\downarrow$ is the sum of probabilities to find $+1/2$
and  $-1/2$ helicities for quarks; while $\delta
f_{q/q'M}=f_{q/q'M}\uparrow -f_{q/q'M}\downarrow$ is the
difference of probabilities to find  $+1/2$ minus  $-1/2$ helicity
for quarks; the quarks being emitted from a parent quark with a specific
helicity. For comparison, $\delta f_{q/q'\pi}(x_\pi)$, $\delta
f_{q/q'K}(x_K)$ and $\delta f_{q/q'\eta}(x_\eta)$ are plotted in Fig.~{\ref{figspli}}.
The polarized quark densities inside the proton can then be obtained using  the following
relations \cite{weise}:
\begin{eqnarray}
\delta u(x) & = & Z\delta u_0(x) + \delta f_{d/u\pi}\otimes\delta
d_0 +
 \delta u_{val}^{\pi}\otimes \delta f_{u/d\pi}\otimes\delta u_0 + \frac{1}{2}\delta f_{u/u\pi}\otimes\delta u_0 \nonumber\\
&& + \frac{1}{4}\delta u_{val}^{\pi}\otimes \delta
f_{\pi}\otimes(\delta u_0+\delta d_0)
+ \delta u_{val}^K\otimes \delta f_K\otimes\delta u_0 + \frac{1}{6}\delta f_{\eta}\otimes\delta u_0 \nonumber\\
&& + \frac{1}{36}\delta u_{val}^{\eta}\otimes \delta f_{\eta}\otimes(\delta u_0+\delta d_0),\nonumber\\
\nonumber\\
\delta d(x) & = & Z\delta d_0(x) + \delta f_{\pi}\otimes\delta u_0
+ \delta d_{val}^{\pi}\otimes \delta f_{\pi}\otimes\delta d_0 +
 \frac{1}{2}\delta f_{\pi}\otimes\delta d_0\nonumber\\
&& + \frac{1}{4}\delta d_{val}^{\pi}\otimes \delta
f_{\pi}\otimes(\delta u_0+\delta d_0) + \delta d_{val}^K\otimes
\delta f_K\otimes\delta d_0 +
 \frac{1}{6}\delta f_{\eta}\otimes\delta d_0\nonumber\\
&& + \frac{1}{36}\delta d_{val}^{\eta}\otimes \delta
f_{\eta}\otimes(\delta u_0+\delta d_0). \label{deltas}
\end{eqnarray}
where $\delta f_{d/u\pi}=\delta f_{u/d\pi}=\cdots=\delta f_{\pi}$
(and so on) and are defined by
Eq.~{(\ref{polspli})} as polarized splitting functions. In Eq.~{(\ref {deltas})}, $\delta u_0$ and
$\delta d_0$ denote  the bare quark distributions inside the
proton, and $\delta u^{\pi}_{val}$, $\delta d^{\pi}_{val}$ and so on
are the polarized quark distributions of mesons, relating to the
cloud which surrounds the bare quarks. The `$\otimes$' symbol
corresponds to the convolution integral, which is defined as:
\begin{eqnarray}
\quad\ \ p\otimes q & = & \int_x^1 \frac{dy}{y} p(y)q(\frac{x}{y}),\nonumber\\
p\otimes q\otimes r & = & \int_x^1 \frac{dy}{y} \int_y^1
\frac{dy'}{y'} p(y')q(\frac{y}{y'})r(\frac{x}{y}).
\end{eqnarray}
Note that the Mellin transform of these equations causes all
$\otimes$ products to convert to ordinary products. $Z$ is the
renormalization constant and in the polarized case it should be
defined by (see Eq.~{(8)} of Ref.~\cite{weise}):
\begin{equation}
Z=H-\frac{3}{2}\Delta f_{\pi}-\Delta f_{K}-\frac{1}{6}\Delta
f_{\eta},\label{Z}
\end{equation}
where $\Delta$ is defined as the first Mellin moment of the
splitting functions $\delta f_{q/q'M}$. In the unpolarized
formulation, $H$ can be determined using the momentum and number sum
rules and equals $1$. In the polarized case, it can be determined
using the Jaffe-Ellis sum rule \cite{ellis}:
\begin{equation}
\int_0^1 xg_1^p(x) dx=0.185\pm 0.010,\label{jaffe}
\end{equation}
where the polarized proton structure function, $g_1^p(x)$, in the
leading order (LO) approximation is given by:
\begin{equation}
g_1^p(x)={1\over 2}\left( {4\over 9}\delta u_{val}(x)+{1\over
9}\delta d_{val}(x) \right).\label{g1p}
\end{equation}
We can substitute $\delta u_{val}(x)$ and $\delta d_{val}(x)$
 with the values given by Eqs.~{(11,15)} from Ref.~\cite{MKY-2011}, which involve $H$,  into Eq.~{(\ref{g1p})} and the result into Eq.~{(\ref{jaffe})} to find the $H$ value. We get the numerical value $H=0.909$.
Substituting this value into Eq.~{(\ref{Z})} yields $Z=0.987$.

In further steps of our calculations, we need to use the sea quark distributions in the chiral quark model, which are given by \cite{weise}:
\begin{eqnarray}
\delta \bar{u}(x) & = & \delta u_{val}^{\pi}\otimes \delta f_{\pi}\otimes\delta d_0 + \frac{1}{4}\delta u_{val}^{\pi}\otimes \delta f_{\pi}\otimes(\delta u_0+\delta d_0)\nonumber\\
&& + \frac{1}{36}\delta u_{val}^{\eta}\otimes \delta f_{\eta}\otimes(\delta u_0+\delta d_0),\nonumber\\
\delta \bar{d}(x) & = & \delta d_{val}^{\pi}\otimes \delta f_{\pi}\otimes\delta u_0 + \frac{1}{4}\delta d_{val}^{\pi}\otimes \delta f_{\pi}\otimes(\delta u_0+\delta d_0)\nonumber\\
&& + \frac{1}{36}\delta d_{val}^{\eta}\otimes \delta f_{\eta}\otimes(\delta u_0+\delta d_0),\nonumber\\
\delta s(x) & = & \delta f_K\otimes(\delta u_0+\delta d_0) + \frac{4}{9}\delta s_{val}^{\eta}\otimes \delta f_{\eta}\otimes(\delta u_0+\delta d_0)\nonumber\\
\delta \bar{s}(x) & = & \delta s_{val}^K\otimes \delta
f_K\otimes(\delta u_0+\delta d_0) + \frac{4}{9}\delta
s_{val}^{\eta}\otimes \delta f_{\eta}\otimes(\delta u_0+\delta
d_0). \label{seas}
\end{eqnarray}

\section{Polarized distribution of the bare quarks}In order to use Eqs.~{(\ref{deltas})} and (\ref{seas})
practically, we need to determine the polarized distributions of
bare quarks inside the proton which, in these equations, are
denoted by $\delta u_0(x)$ and $\delta d_0(x)$.  To extract these
distributions, we do as follows.

We fit two simple functions with $\delta q_{val}^p$ data
 from ~\cite{hermes,smc} and find their ratio, $\delta u_{val}/
\delta d_{val}$, and also suppose that this ratio is unchanged
when $Q^2\rightarrow Q_0^2$. This assumption is reliable because
when we move from $Q_0^2$ to $Q^2$,
 all valence quarks share an equal proportion of their momentum with gluons and sea quarks. Then we can rewrite Eq.~{(\ref{g1p})} as:
\begin{equation}
g_1^p(x)={1\over 2}\left( {4\over 9}{\delta u_{val}\over \delta
d_{val}}\delta d_{val}+{1\over 9}\delta d_{val}
\right).\label{g1p2}
\end{equation}
The only unknown function is then $\delta d_{val}(x)$, which can be
determined by fitting the right hand side of
Eq.~{(\ref{g1p2})} with the available experimental  data for
$g_1^p(x)$ at high energy scales \cite{e143,compass}. We evolve
it down to $Q_0^2$ to find $\delta d_0(x)$. We are also able to
find $\delta u_0(x)$ from the known ratio $\delta u_{val}/
\delta d_{val}$. The final results are given by:
\begin{eqnarray}
x\delta u_0(x)&=&2.313\ x^{1.100}\ (1-x)^{1.908},\nonumber\\
x\delta d_0(x)&=&-0.852\ x^{0.964}\ (1-x)^{2.485}. \label{bare}
\end{eqnarray}

The only functions that remain unknown in the rest of our
calculations are the PPDFs of the mesons, denoted by $\delta
q^M_{val}$ in Eqs.~{(\ref{deltas})} and (\ref{seas}). We  use the
following strategy to find the polarized valence densities in all
mesons. We first consider the following functions  for the
polarized valence distribution in mesons:
\begin{equation}
\delta q_{val}^M=a\; x^b\; (1-x)^c\; P_M(x), \label{qpi}
\end{equation}
where the superscript $M$ denotes meson and for the pion we have
$P_{\pi}(x)=1$. To calculate the other functions $P_K(x)$ and
$P_{\eta}(x)$ for the kaon and eta, we need to resort to a  method which will be  explained
in the following subsections.

\subsection{Monte-carlo simulation -- Meson polarized quark densities}Meson polarized distributions for kaon and eta will be specified
if we determine $P_K(x)$ and $P_{\eta}(x)$ in Eq.~{(\ref{qpi})}.
The quantity $P_K(x)$ can be determined using Monte Carlo
(MC) simulation and $P_{\eta}(x)$ can be obtained by expanding the
meson PDFs as a function of quark mass.

In the MC simulation which we introduce, $P_K(x)=\delta
q_{val}^K/\delta q_{val}^{\pi}$ should be found. This ratio is
related to the unpolarized  $q_{val}^K/q_{val}^{\pi}$
data~\cite{badier}. Two distinct MC algorithms are employed. In one
of them we use the unpolarized values and their errors directly
and in the other algorithm, these values are used  as controlling
conditional parameters to generate random numbers to calculate and
estimate the polarized values. The first algorithm  can be
expressed as follows.
There are experimental data  for the ratio~\cite{badier}:
\begin{equation}
{q_{val}^K \over q_{val}^{\pi}}={q_{val}^K\uparrow
+q_{val}^K\downarrow \over q_{val}^{\pi}\uparrow +
q_{val}^{\pi}\downarrow}=r\pm \delta r.\label{datrat}
\end{equation}
Experimental data for the unpolarized  pion  valence distributions
are also available~\cite{conway}:
\begin{equation}
q_{val}^{\pi}\uparrow +q_{val}^{\pi}\downarrow =d\pm \delta
d.\label{datpi}
\end{equation}
We can therefore find that $q_{val}^K\uparrow +q_{val}^K\downarrow =r.d\pm
\delta (r.d)$ where $\delta (r.d)=d.\delta r+r.\delta d$ (we
ignore the product term $\delta r.\delta d$). On the other hand,
we know that the difference of two numbers between $0$ and $1$
should lie between $-1$ and their sum, hence:
\begin{equation}
-1\leq \delta q_{val}^{\pi}\leq n\, q_{val}^{\pi},\label{les}
\end{equation}
where  we have inserted $n$ in Eq.~{(\ref{les})} for some other
possible theoretical and/or experimental considerations.
Nevertheless  we take $n=1$ in our calculations. From
Eq.~{(\ref{les})} we have:
\begin{equation}
-1\leq q_{val}^{\pi}\uparrow -q_{val}^{\pi}\downarrow \leq
n\;(q_{val}^{\pi}\uparrow +q_{val}^{\pi}\downarrow).
\end{equation}
From Eq.~{(\ref{datpi})} and the upper limit of its right hand
side we get:
\begin{eqnarray}
&& -1 \leq d\pm (\delta d) -2q_{val}^{\pi}\downarrow \leq n(d+\delta d),\nonumber\\
\nonumber\\
& \Rightarrow & {1+d \over 2} \geq q_{val}^{\pi}\downarrow \mp
\left( {\delta d \over 2}\right) \geq -{n\delta d+(n-1)d \over
2}.\qquad
\end{eqnarray}
We can generate random numbers between these two limits and find
$q_{val}^{\pi}\downarrow$ while the uncertainty  $\delta d$ is
known. The same method can be used to find $q_{val}^K\downarrow$
and hence we can calculate $\delta q_{val}^K/\delta
q_{val}^{\pi}=(r.d-2q_{val}^K\downarrow)/(d-2q_{val}^{\pi}\downarrow)$
where we  use Eqs.~{(\ref{datrat}) and (\ref{datpi})}. As a result
$P_K(x)$ is calculated. One can determine the maximum value of $n$
using the theoretical and experimental values of $\delta
q_{val}^M/q_{val}^M$. This could be an issue for further
research activity.

The second method is based on direct generation of random values
between $0$ and $1$ including all $q_{val}^M\uparrow$ and
$q_{val}^M\downarrow$ for both pion and kaon, and eliminating the
results of $q_{val}^{\pi}$ which lie outside the interval $d\pm \delta d$
 (Eq.~{(\ref{datpi})}) and those of $q_{val}^K$ which lie outside
the interval $r.d\pm \delta (r.d)$. The difference between the outputs of
these two algorithms, plus the uncertainty that we mentioned
before, can be used to calculate the error of the calculations.

\begin{figure}[htp]
\begin{center}
\includegraphics[width=8.1 cm]{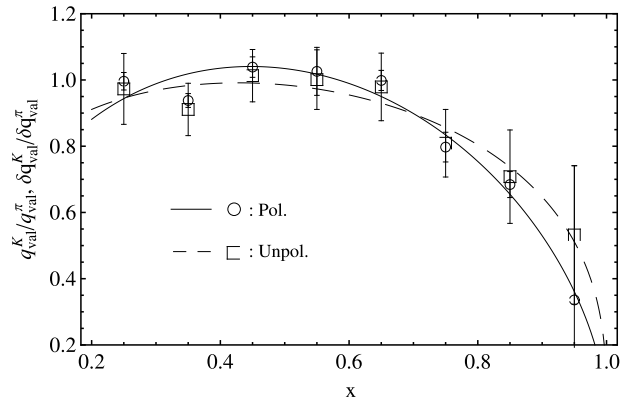}
{\caption{The unpolarized data\protect~\cite{badier} and the
polarized Monte-Carlo results for the ratio $\delta q_{val}^K/\delta
q_{val}^{\pi}$. \label{figrat}}}
\end{center}
\end{figure}

Although the results of MC algorithms always depend on the
running duration of the program (because of their probabilistic
structure), we increase the number of random values in order that
the results vary less than 10 percent in two consecutive runs
of the program. The unpolarized data and polarized values produced
by MC simulation for the ratio concerned, together with the
functions which have been fitted to them, are depicted in
Fig.~{\ref{figrat}}; consequently we get:
\begin{equation}
P_K(x)=2.170\ x^{0.478}\ (1-x)^{0.591}. \label{pK}
\end{equation}
By obtaining $P_{K}(x)$, we are able to determine $P_{\eta}(x)$
and finally to extract the valence quark distribution of the eta meson,
which we explain in the following subsection.
\subsection{Valence quark distribution for eta meson}
To find $P_{\eta}(x)$ we need to have the quark distribution functions depend on their masses. If we consider the mass of the quarks as a factor that causes  their densities to be different, we can obtain:
\begin{eqnarray}
\delta q_{val}^{\pi}&=&f(m,\cdots)|_{m=m_l}\times f(m,\cdots)|_{m=m_l},\nonumber\\
\delta q_{val}^K&=&f(m,\cdots)|_{m=m_l}\times f(m,\cdots)|_{m=m_s},\nonumber\\
\delta q_{val}^{\eta}&=&f(m,\cdots)|_{m=m_s}\times
f(m,\cdots)|_{m=m_s}, \label{qmass}
\end{eqnarray}
where $f$ could be any function of quark mass, $m$, and all other
QCD parameters ($x$, $Q_0$, $\Lambda_{QCD}$, $\cdots$). The
concealed logic in Eq.(\ref{qmass}) is that the pion contains two
light quarks, the kaon contains one light and one strange quark,
and so forth. The product of two $f$ functions in
Eq.~{(\ref{qmass})} is justifiable by  the probabilistic nature of
the distributions. The mass of light quarks is denoted by
$m_l=m_u\simeq m_d$ and $m_s$ is the mass of strange quark. The
expansion of Eq.~{(\ref{qmass})} yields:
\begin{eqnarray}
\delta q_{val}^{\pi} & = & \left(f|_{m=0}+m_l{\partial f\over\partial m}|_{m=0}+\mathcal{O}(m^2)\right)\left(f|_{m=0}+m_l{\partial f\over\partial m}|_{m=0}+\mathcal{O}(m^2)\right),\nonumber\\
& = & f_0^2+2m_lf_0f'_0+\mathcal{O}(m^2),
\end{eqnarray}
where $f_0=f|_{m=0}$ and $f'_0=(\partial f/\partial m)|_{m=0}$.
Doing the same calculations for the kaon and eta meson we find:
\begin{eqnarray}
\delta q_{val}^K & = & \left(f_0+m_sf'_0+\mathcal{O}(m^2)\right)\left(f_0+m_lf'_0+\mathcal{O}(m^2)\right)\nonumber\\
& = & f_0^2+(m_s+m_l)f_0f'_0+\mathcal{O}(m^2),\\
\nonumber\\
\delta q_{val}^{\eta} & = & f_0^2+2m_sf_0f'_0+\mathcal{O}(m^2).
\end{eqnarray}
Defining $D=f'_0/f_0$ we have:
\begin{eqnarray}
{\delta q_{val}^K \over \delta q_{val}^{\pi}} & = & P_K(x) = {f_0^2+(m_s+m_l)f_0f'_0 \over f_0^2+2m_lf_0f'_0} = {1+(m_s+m_l)D \over 1+2m_lD}, \label{pkopp}\\
\nonumber\\
{\delta q_{val}^{\eta} \over \delta q_{val}^{\pi}} & = &
P_{\eta}(x) = {f_0^2+2m_sf_0f'_0 \over f_0^2+2m_lf_0f'_0} =
{1+2m_sD \over 1+2m_lD}. \label{peta}
\end{eqnarray}
Consequently, by writing $D$ in terms of $P_K$ from
Eq.~{(\ref{pkopp})} and substituting it into Eq.~{(\ref{peta})} we
find:
\begin{equation}
P_{\eta}(x)=2P_K(x)-1.\label{pepk}
\end{equation}
This equation allows us to determine the valence quark
distributions of the eta meson, principally. Now we should determine
the unknown parameters ($a$, $b$ and $c$) in Eq.~{(\ref{qpi})}.

\begin{figure}[htp]
\begin{center}
\includegraphics[width=10 cm]{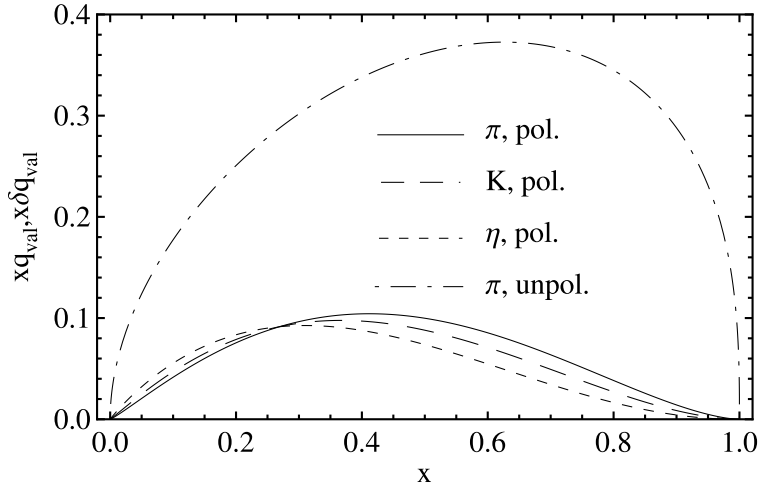}
{\caption{The polarized valence densities in mesons. A
corresponding unpolarized PDF \protect\cite{piongrv1,piongrv2} is
also included for comparison. \label{mesons}}}
\end{center}
\end{figure}

Substituting Eqs.~{(\ref{pepk})} and (\ref{pK}) into
Eq.~{(\ref{qpi})} and the obtained result into
Eqs.~{(\ref{deltas},\ref{seas})} and then fitting the right hand
side of this equation with the experimental data for proton parton
distributions (in the polarized case, their large errors have made
them unusable for fitting processes) or the results of
phenomenological collaborations, we can find parameters $a$, $b$ and $c$
from Eq.~{(\ref{qpi})}. We choose the average of
GRSV \cite{GRSV,grv,grsv} and AAC \cite{goto} for fitting. Due to
the slightly imprecise results for $Q^2<4\ GeV^2$, we excluded the
BB model from our calculations . The results are:
\begin{eqnarray}
a & = & 1.100\pm 0.235,\nonumber\\
b & = & 0.686\pm 0.118,\nonumber\\
c & = & 1.073\pm 0.274.\label{parmeson}
\end{eqnarray}
Taking these parameters into account and using the results of
Section 3.1 of Ref.~\cite{kshvzn}, evolution of the PDFs based on
the constituent valon model (for mesons) is straightforward
\cite{kshvzn,MK}. The results for the polarized valence distribution
functions at $Q^2=3\ \mathrm{GeV}^2$ are depicted in
Fig.~{\ref{mesons}} and compared with unpolarized distribution. Other meson
densities, extracted  from the valon model, are listed in the
Appendix.

As an adjunct to this study and to complete the discussion, let us
review how to get the numerical values which are listed in the
appendix. In analogue to Eq.~{(33)} from Ref.~\cite{kshvzn}, but for the polarized case, we can write:
\begin{equation}
\delta M_{val}(n,Q^2)=\delta V(n)\times \delta
M_{NS}(n,Q^2),\label{Mval}
\end{equation}
where $\delta V(n)$ is the moment of the polarized valon
distributions. Note that in the unpolarized case we had two
valons with corresponding distributions which were generally
different. In Appendix A of Ref.~\cite{kshvzn}, we have shown
that their difference can be obtained using the number and
momentum sum rules. For the pion -- which consists of two light
valence quarks -- the calculations of Ref.~\cite{kshvzn} showed us
that we can take their valon distributions to be equal to each other.  But in the polarized
case, the lack of sufficient  theoretical sum rules force us to
suppose that all valence distributions inside each meson are
equal. Hence, instead of $V_1(n)$ and $V_2(n)$ from Eq.~{(33)}
of Ref.~\cite{kshvzn} we have only $\delta V(n)$, which is assumed
to have the following form:
\begin{equation}
\delta V(n)={B(p+n,q+1)\over B(p+1,q+1)},
\end{equation}
where $B$ is the Euler beta function and $p$ and $q$ are two
(valon) free parameters. There are also two (QCD) free parameters,
\ie $Q_0$ and $\Lambda_{QCD}$, which exist in the definition
of $\delta M_{NS}(n,Q^2)$. These four parameters can be
determined by fitting the right hand side of
Eq.~{(\ref{Mval})} to Eq.~{(\ref{qpi})}, using Eqs.~{(\ref{pK}),}
(\ref{pepk}) and (\ref{parmeson}) for mesons. Having these four
parameters, we can then calculate the distributions of $\delta
M_{\Sigma}(n,Q^2)$ (the moment of singlet sector of the
distributions) and $\delta M_g(n,Q^2)$ (the moment of the gluon
distribution):
\begin{eqnarray}
\delta M_{\Sigma}(n,Q^2) & = & 2\delta V(n)\times \delta M_S(n,Q^2),\label{sig}\\
\delta M_g(n,Q^2) & = & 2\delta V(n)\times \delta
M_{qg}(n,Q^2).\label{gel}
\end{eqnarray}
To find the required distribution functions ($\delta
\Sigma(x,Q^2)$ and $\delta g(x,Q^2)$), we  can use the inverse
Mellin transform (Eq.~{(34)} from Ref.~\cite{kshvzn}):
\begin{equation}
xq(x,Q^2)={1\over 2\pi
\mathrm{i}}\int_{c-\mathrm{i}\infty}^{c+\mathrm{i}\infty}{dx\over
x^{n-1}}M(n,Q^2)\;.\label{inv}
\end{equation}
A numerical method to cope with the integral in Eq.(\ref{inv}) has
been introduced in Ref.~\cite{MK}. In all of the numerical methods, we
simply fit the moment of a definite function (which contains free
parameters) to the experimental data or the results  of a known
function -- for example, Eqs.~{(\ref{Mval}),} (\ref{sig}) or
(\ref{gel}) -- to find the free parameters. In our fitting
procedure, we assume the unknown function to be  $a\, x^b (1-x)^c$
with free parameters  $a$, $b$ and $c$. Because
Eqs.~{(\ref{Mval}),} (\ref{sig}) and (\ref{gel}) depend on $Q^2$,
the free parameters ($a$, $b$ and $c$) take  different values at
each energy scale. We compute them for a wide range of energies ($Q^2=0.7\
\mathrm{to}\ 100\ \mathrm{GeV}^2$) and categorize them in the
appendix for any possible practical usage.

\section{Parton orbital angular momentum}
Since we now have the polarized parton densities for mesons, we can
investigate the first Mellin moment of the singlet, non-singlet
and gluon sectors of the meson and finally determine its spin. To
avoid spin crises, we need to calculate the gluon and quark
orbital momenta. Their analytical calculations are fully discussed
in Ref.~\cite{ji}. The leading-log evolution of the quark and gluon
orbital angular momenta are:
\begin{eqnarray}
{d\over dt} \left( \begin{array}{c} L_q\\L_g \end{array} \right) &
= &
{\alpha_s(t)\over 2\pi} \left( \begin{array}{cc} -{4\over 3}C_F & {n_f\over 3}\\
{4\over 3}C_F & -{n_f\over 3}\end{array} \right) \left(
\begin{array} {c} L_q\\L_g \end{array} \right) + {\alpha_s(t)\over
2\pi} \left( \begin{array}{cc} -{2\over 3}C_F & {n_f\over 3}\\
-{5\over 6}C_F & -{11\over 2} \end{array} \right) \left(
\begin{array}{c} \Delta\Sigma\\ \Delta g \end{array}
\right)\;,\label{angular}
\end{eqnarray}
where $L_q$ and $L_g$ are the orbital angular momentum of quarks
and gluons respectively; $t=\ln(Q^2/\Lambda_{QCD}^2)$; $C_F=4/3$
and $\Delta \Sigma$ and $\Delta g$ are the first Mellin  moment of
the distributions $\delta\Sigma$ and $\delta g$, \ie:
\begin{equation}
\left( \begin{array}{c} \Delta\Sigma\\ \Delta g \end{array}\right) = \int_0^1\left( \begin{array}{c} \delta\Sigma(x)\\
\delta g(x) \end{array}\right) dx\;.
\end{equation}
As $Q^2$ increases, $\Delta\Sigma$ decreases very slightly.
It is therefore considered constant in Ref.~\cite{ji} and also in
our calculations. The dependence of $\Delta g$ on $Q^2$
 can be obtained via:
\begin{eqnarray}
\Delta g(t) & = & -{4\over \beta_0}\Delta\Sigma + {t\over t_0}\left( \Delta g_0 + {4\over \beta_0}\Delta\Sigma \right), \nonumber\\
\Delta\Sigma & = & \mathrm{const}, \label{delg}
\end{eqnarray}
where $t_0=t(Q_0^2)$ and  the first universal coefficient of the QCD
$\beta$-function is  $\beta_0=11-2n_f/3$. If we solve
Eq.~{(\ref{angular})} for a meson and use the following boundary
condition:
\begin{equation} 0={1\over 2}\Delta\Sigma + \Delta g(0) +
L(0),\label{mesbound}
\end{equation}
we will get \cite{MKY-2011}:
\begin{eqnarray}
L_q(t) & = & -{1\over 2}\Delta\Sigma + \left(t/t_0\right)^{-2(16+3n_f)/(9\beta_0)}(L_q(0) + {1\over 2}\Delta\Sigma ), \nonumber\\
L_g(t) & = & -\Delta g(t) +
\left(t/t_0\right)^{-2(16+3n_f)/(9\beta_0)} ( L_g(0) + \Delta
g(0)).\label{lqlg}
\end{eqnarray}
By summing up the two equations in {(\ref{lqlg})}, the total orbital angular momentum is obtained:
\begin{equation}
L(t)=L_q(t)+L_g(t).\label{l}
\end{equation}
\begin{figure}[htp]
\begin{center}
\includegraphics[width=9 cm]{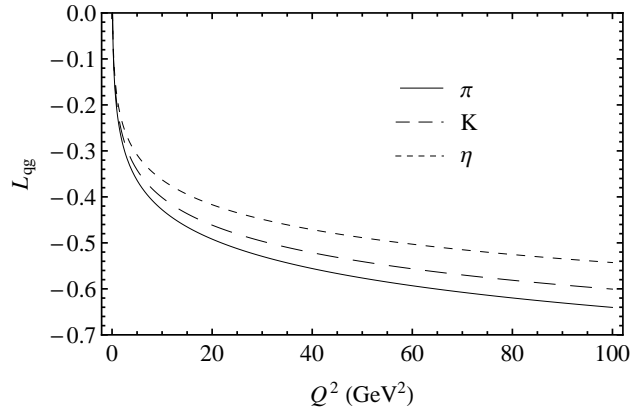}
\caption{The orbital angular momentum $L(t)$ for mesons with
respect to $Q^2$.\label{figang}}
\end{center}
\end{figure}
The initial value of parton angular momentum for a meson, \ie
$L(0)=L_q(0)+L_g(0)$, can be obtained using the first Mellin moment
of the total quark and gluon helicity distributions and their
orbital angular momentum respectively, \ie Eq.~{(\ref{mesbound})}.
Knowing $L(0)$ and using Eqs.~{(\ref{lqlg})} and (\ref{l}) we can
calculate $L(t)$ at all energy ranges. The results for mesons
are shown in Fig.~{\ref{figang}}. They are in good agreement with
those in Ref.~\cite{MKY-2011}, in which another aspect of $\chi QM$
was used. This agreement confirms the validity of our
recent calculations. The total spin of hadrons at all energies can
be obtained from:
\begin{equation}
S(t)=\frac{1}{2}\Delta \Sigma+ \Delta g(t)+
L_{qg}(t)\;,\label{spin}
\end{equation}
which easily leads to a value of zero, using Eqs.~{(\ref{delg})} and
(\ref{lqlg}), as expected.
\section{ Mass dependence of the proton quark distribution -- evolution of sea quark densities}
\begin{figure}[htp]
\begin{center}
\includegraphics[width=8 cm]{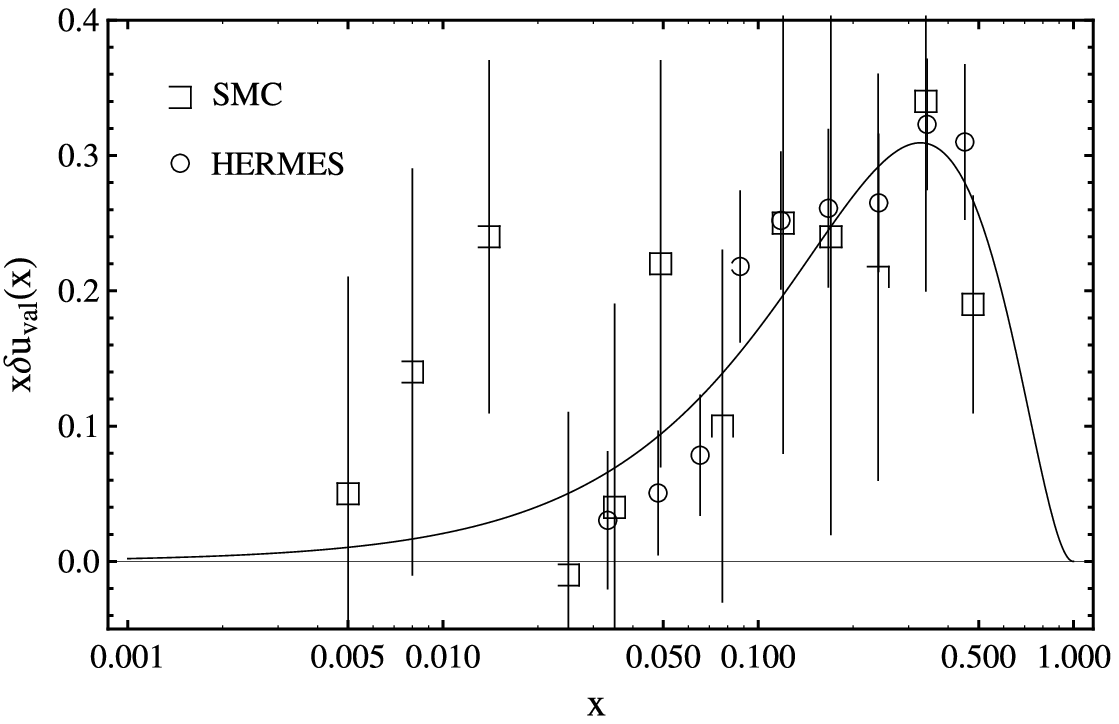}
\includegraphics[width=8 cm]{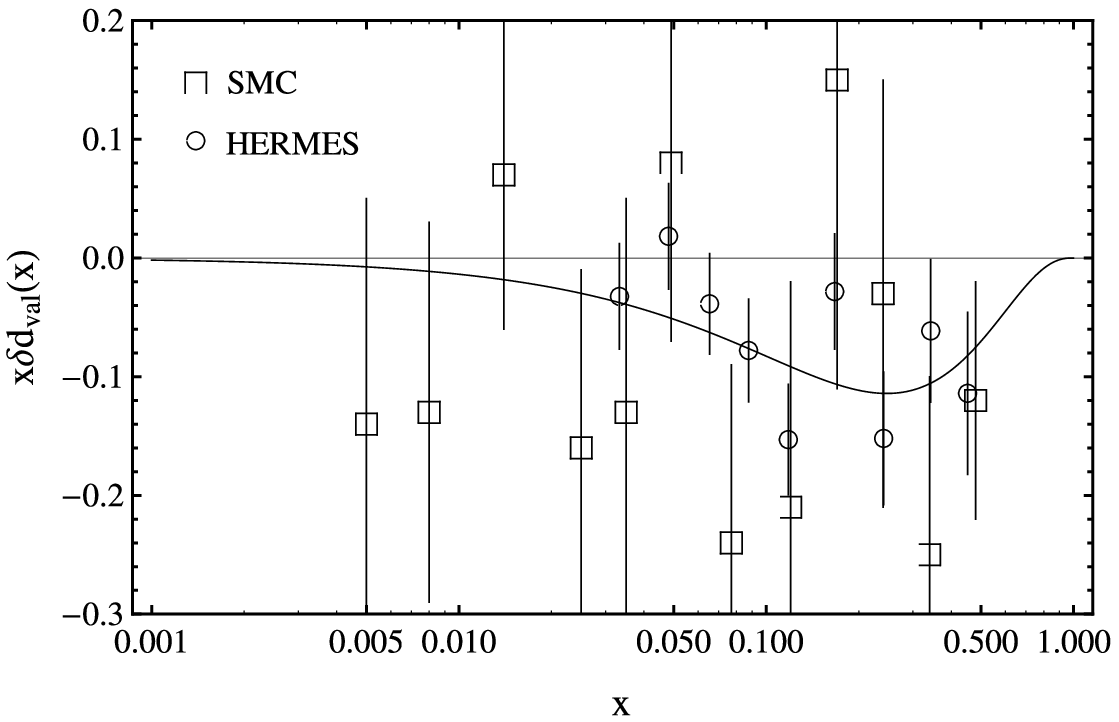}
\caption{The polarized valence densities inside a proton at $Q^2=3\
\mathrm{GeV}^2$. Note that HERMES data \protect\cite{hermes} are
at $Q^2=2.5\ \mathrm{GeV}^2$ and the SMC data \protect\cite{smc}
are at $Q^2=3\ \mathrm{GeV}^2$.\label{valsmc}}
\end{center}
\end{figure}

Accessing the bare quark distributions inside the proton, using
 Eq.~{(\ref{bare})} and the valence densities of mesons with
Eq.~{(\ref{qpi})}, we can obtain the polarized quark
distribution inside the proton using Eqs.~{(\ref{deltas})} and
(\ref{seas}). To evolve the valance density to high energies, we
use the evolution of non-singlet moments,  based on the  the DGLAP equations.
The required relation to evolve the non-singlet moments is as
follows:
\begin{equation}
\delta M_{val}(n,Q^2)=\delta M_{val}(n,Q_0^2)\times \delta
M_{NS}(n,Q^2)\;.\label{val}
\end{equation}
The term $\delta M_{NS}(n,Q^2)$ is available from QCD calculations
and $\delta M_{val}(n,Q_0^2)$ is the Mellin moment of the valence
distribution, which has been previously obtained based on the $\chi
QM$ at low $Q^2$ (see Eq.(\ref{deltas}) and Eq.(\ref{seas})). The
evolved valence densities inside the proton are indicated in
Fig.~{\ref{valsmc}} and compared with available experimental data.

The evolution of sea quark densities inside the proton is not as
simple as that of valence quarks. In this case we need the singlet moment
which does not relate individually to the moment of sea quarks but
relates to the summation of all quark moments $(\delta\Sigma)$.
Although there are some methods that solve this
difficulty~\cite{GRSV}, we are looking for a different method by
applying  the mass of the quarks in the calculations. We first
assume that all four light sea quarks are eigenstates of the evolution
operator:
\begin{eqnarray}
|\delta q_{sea};Q^2\rangle & = & \mathrm{\mathbf{E}}|\delta
q_{sea};Q_0^2\rangle \nonumber\\
\Rightarrow |\delta q_{sea};Q^2\rangle & = & E_q(x,Q^2)|\delta
q_{sea};Q_0^2\rangle, \label{ket}
\end{eqnarray}
in which \textrm{\textbf{E}} is the evolution operator and $E_q$
is its eigenvalue. Two distinct cases can occur:

\vspace*{3 mm} {\bf 1. Degenerate state}
\vspace*{3 mm}\\
In this case we have:
\begin{equation}
E_{\bar{u}}=E_{\bar{d}}=E_s=E_{\bar{s}}=E(x,Q^2),
\label{degenerate}
\end{equation}
and subsequently:
\begin{equation}
\delta q_{sea} (x,Q^2)=E(x,Q^2)\times \delta q_{sea} (x,Q_0^2).
\label{operator}
\end{equation}
By summing both sides of Eq.~{(\ref{operator})} for the four
light quarks and then factorizing $E(x,Q^2)$, we will finally
reach the following relation:
\begin{eqnarray}
E(x,Q^2) & = & \frac{[\delta \bar u+\delta \bar d+\delta s+\delta
\bar s](x,Q^2)} {[\delta \bar u +\delta \bar d+\delta s+\delta
\bar s](x,Q_0^2)} = \frac{[\delta \Sigma-\delta u_{val}-\delta
d_{val}] (x,Q^2)}{[\delta \bar u +\delta \bar d+\delta s+\delta
\bar s](x,Q_0^2)}. \label{operatorn}
\end{eqnarray}
In the second fraction of Eq.~{(\ref{operatorn})}, the evolved
valence densities can be obtained from Eq.~{(\ref{val})} and the
evolved distributions for $\delta \Sigma$ can be obtained, using the
notation of Ref.~\cite{hwa1}, as:
\begin{eqnarray}
&& \delta M_{\Sigma}(n,Q^2) = \left[\delta Mu_{val}(n,Q_0^2)+
\delta Md_{val}(n,Q_0^2)\right]\times \delta M_s(n,Q^2).\qquad
\label{singlet}
\end{eqnarray}
The denominator in Eq.~{(\ref{operatorn})} can be obtained from
 $\chi QM$ (see Eq.~(\ref{seas})). Having the functional
form of $E(x,Q^2)$, the evolution of individual sea quark
densities will be possible. This is also the method which is used
in~\cite{MKY-2011} based on another aspect of $\chi QM$.

\vspace*{3 mm} {\bf 2. Non-degenerate state}
\vspace{3 mm}\\
In this case, we cannot factorize $E_q$ in
Eq.~{(\ref{operator})}. Assuming that the eigenvalues in this
equation depend on $Q^2$ through the running mass of the quarks,
we can write:
\begin{eqnarray}
\left[\delta \Sigma -\delta u_{val}-\delta d_{val}\right](x,Q^2) & = & E_u(x,m_u(Q^2))\times\delta \bar{u}(x,Q_0^2) \nonumber\\
& + & E_d(x,m_d(Q^2))\times\delta \bar{d}(x,Q_0^2) \nonumber\\
& + & E_s(x,m_s(Q^2))\times[\delta s(x,Q_0^2)+\delta
\bar{s}(x,Q_0^2)], \label{nondeg}
\end{eqnarray}
where $m_q(Q^2)=m_{\bar{q}}(Q^2)$ for all quarks and due to the
equality of the $s$ and $\bar{s}$ quarks masses, the two
eigenvalues of the strange distributions (in the last bracket) are
equal.

In the modified minimal subtraction ($\overline {MS}$) scheme, the
renormalization group equation for the running mass of quarks has
the following form~\cite{wscheme,Stir-ellis}:
\begin{eqnarray}
&& \hspace*{-5 mm} \left[ Q^2{\partial \over\partial
Q^2}-\beta(\alpha_s){\partial \over\partial \alpha_s}+({1\over 2}+
\gamma_m(\alpha_s))m{\partial \over\partial m}
\right]R(Q^2/\mu^2,\alpha_s,m/Q)=0.
\end{eqnarray}
The running mass equation $m(Q^2)$ (in analogy with the running
coupling constant) is governed by:
\begin{equation}
Q^2{\partial m \over\partial Q^2}=-\gamma_m(\alpha_s)m(Q^2),
\end{equation}
and finally its solution is:
\begin{equation}
m(Q^2)=m(\mu^2)\exp \left[ -\int_{\mu^2}^{Q^2}{d Q^2 \over
Q^2}\gamma_m (\alpha_s(Q^2)) \right].
\end{equation}
The numerical values for the light quark masses, which are denoted
here by $m(\mu^2)$, are those which were indicated just below
Eq.(\ref{mm}).

We assume the following function  for the $E_q$s in
Eq.~{(\ref{nondeg})}:
\begin{equation}
E_q=A_q(m_q)\ x^{B_q(m_q)}\ (1-x)^{C_q(m_q)}. \label{Eq}
\end{equation}
Eq.~{(\ref{operator})} shows that $E(x,Q^2)$ should be equal to
$1$ at $Q^2=Q_0^2$, and hence in Eq.~{(\ref{nondeg})}
 $E_u=E_d=E_s\rightarrow 1$ when $Q^2\rightarrow Q_0^2$. However, the natures of the two sides of Eq.~{(\ref{nondeg})}
 are actually different. The left hand side of Eq.~{(\ref{nondeg})} can be calculated, based on the  Valon framework~\cite{hwa4}, while its right hand side
 comes from $\chi QM$~\cite{MKY-2011,kshvzn}. As a result we consider an additional coefficient to fill this gap between the two models and
 write $E_u=E_d=E_s=N$. Thereupon in Eq.~{(\ref{Eq})}, when $Q^2\rightarrow Q_0^2$ we have:
\begin{eqnarray}
\mathrm{as}\ Q^2\rightarrow Q_0^2:&&\nonumber\\
&\lim A_u=\lim A_d=\lim A_s=N,&\nonumber\\
&\lim B_u=\lim B_d=\lim B_s=0,&\nonumber\\
&\lim C_u=\lim C_d=\lim C_s=0,&
\end{eqnarray}
where $N$ can be determined from Eq.~{(\ref{nondeg})} when its
left hand side is calculated at $Q^2=Q_0^2$ using a regression
method~\cite{bates,meyer,ratkowsky}.

Also, we know  that quarks at high energy scales can be considered
massless~\cite{Stir-ellis}. According to Eq.~{(\ref{Eq})}, this
condition implies the following limits:
\begin{eqnarray}
\mathrm{as}\ Q^2\rightarrow \infty :&&\nonumber\\
&\lim A_u=\lim A_d=\lim A_s,&\nonumber\\
&\lim B_u=\lim B_d=\lim B_s,&\nonumber\\
&\lim C_u=\lim C_d=\lim C_s.&
\end{eqnarray}

\begin{figure}[h]
\begin{center}
\includegraphics[width=9.5 cm]{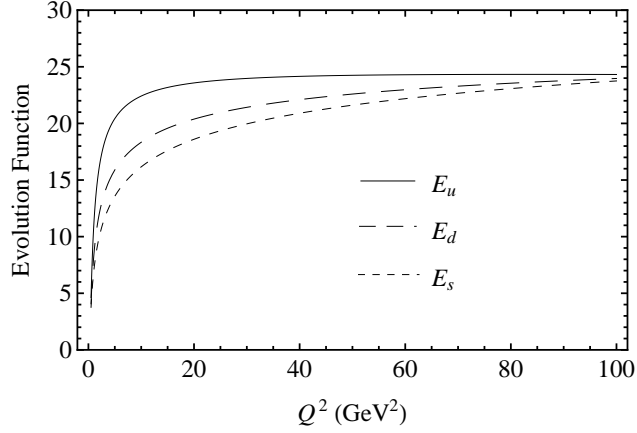}
\caption{The evolution functions in Eq.~{(\ref{nondeg})} at
$x=0.3$.\label{evo}}
\end{center}
\end{figure}

One of the simple functions which satisfies these conditions can be
indicated by:
\begin{eqnarray}
A_q(m_q)&=&N\left( {m_q(Q^2)\over m_q(\mu^2)} \right)^{-A},\nonumber\\
\nonumber\\
\left( \begin{array}{c} B_q(m_q)\\ C_q(m_q)
\end{array}\right)&=&\left( \begin{array}{c} B
\\ C \end{array}\right)\left[-\log \left( {m_q(Q^2)\over m_q(\mu^2)} \right) \right]\left( {\log [m_q(Q^2)]\over \log [m_q(Q'^2)]} \right),
\label{ABCq}
\end{eqnarray}
here $Q'^2$ refers to  the limit of large energy value in which  the
quarks can be considered massless.

Using  Eq.~{(\ref{nondeg})} at $Q^2=Q_0^2$  will tend the value
of $N$ to 2.785. To find $A$, $B$ and $C$, we substitute the
evolved $\delta\bar{u}$, $\delta\bar{d}$ and $\delta\bar{s}$
distributions into the first equation of the DGLAP equations:
\begin{eqnarray}
&& \hspace*{-3 mm} {d\over d\log Q^2}\delta q(x,Q^2)={\alpha_s
\over 2\pi}\int_x^1 {dy\over y}\left[\delta q(y,Q^2)\delta
P_{qq}({x\over y})+\delta g(y,Q^2)\delta P_{qg}({x\over
y})\right].\qquad \label{DGLAP}
\end{eqnarray}
(Note that, according to Ref.~\cite{urek} we can calculate the
`frozen' gluon distribution, $\delta g(x)$, from $\chi$QM and
evolve it using $\delta M_g(n,Q^2)$). Considering
Eq.~(\ref{DGLAP}), we get three equations for $\delta\bar u$,
$\delta\bar d$ and $\delta\bar s$, which should be solved
numerically for $A$, $B$ and $C$. Consequently the results for
the parameters in Eq.(\ref{ABCq}) are:
\begin{equation}
A=0.797,\ B=-0.502,\ C=-1.306. \label{ABC}
\end{equation}

\begin{figure}[htp]
\begin{center}
\includegraphics[width=8 cm]{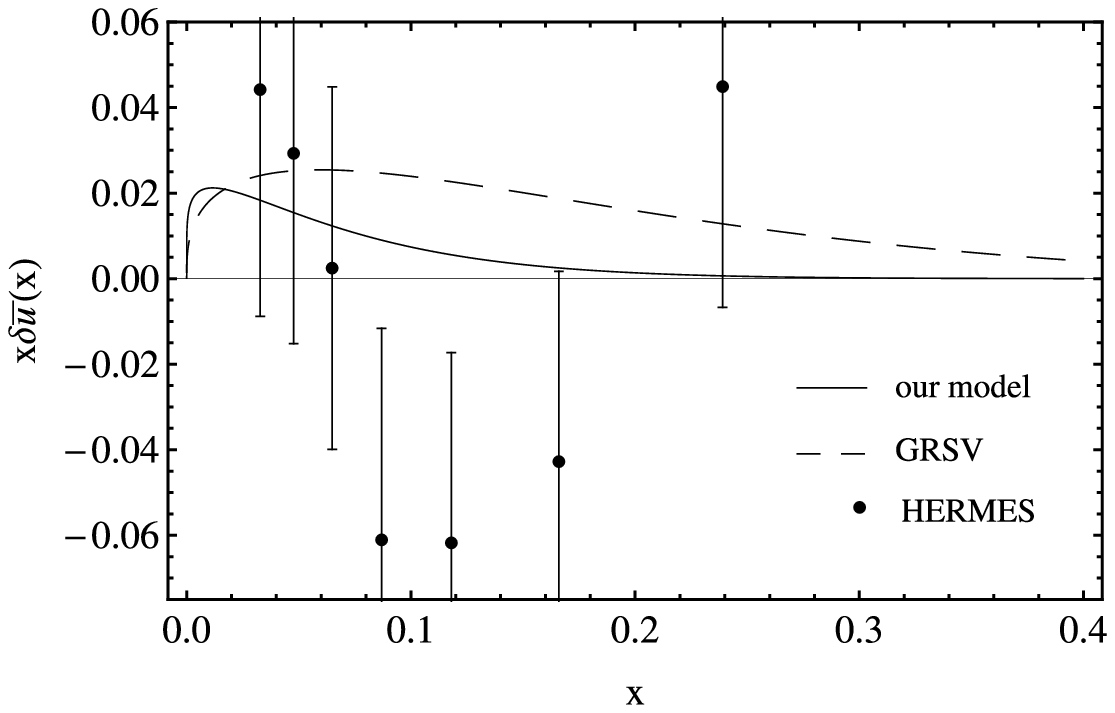}
\includegraphics[width=8 cm]{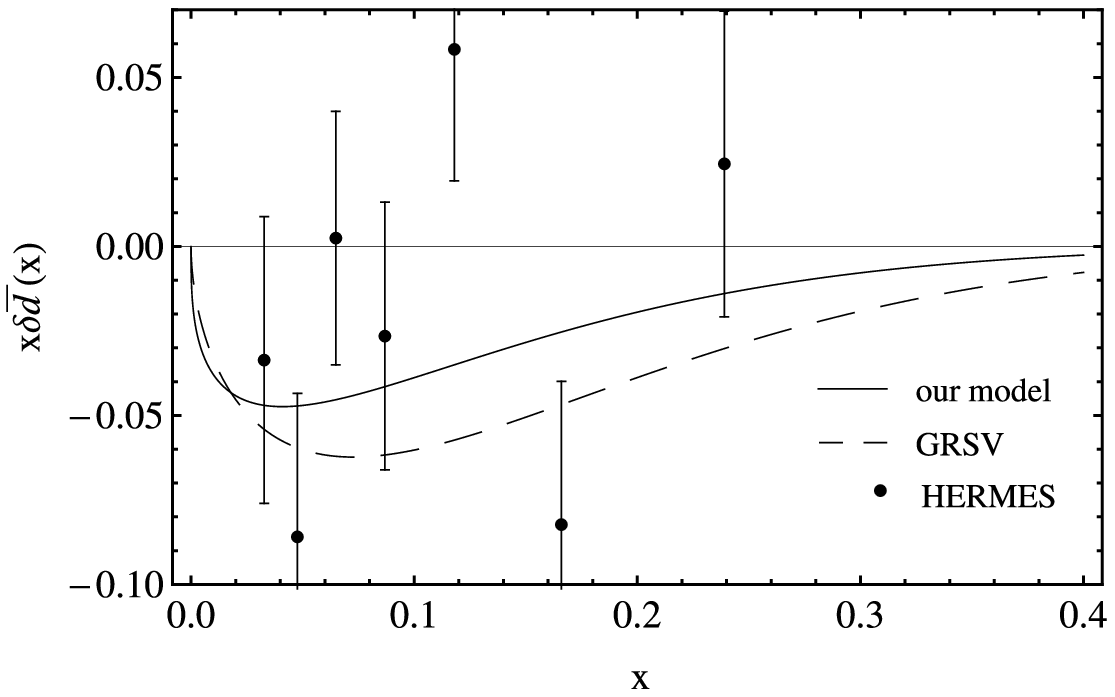}
\includegraphics[width=8 cm]{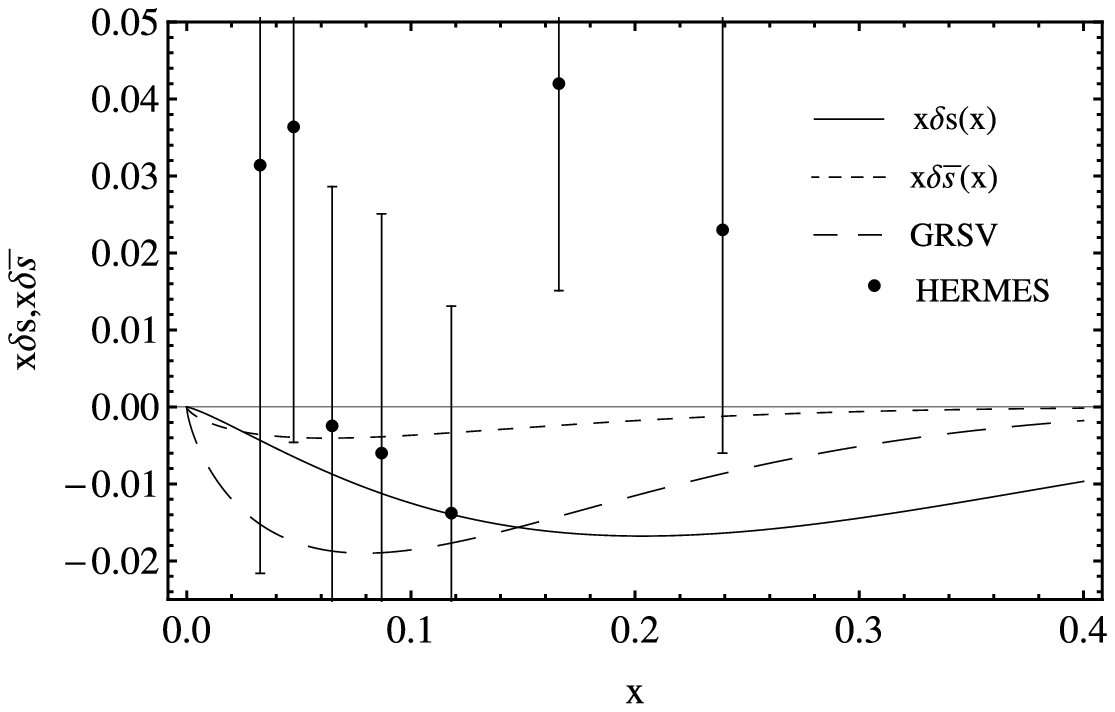}
\caption{The sea quark distributions at $Q^2=3\ \mathrm{GeV}^2$
together with the HERMES data
\protect\cite{hermes}.\label{evosea}}
\end{center}
\end{figure}
Substituting Eqs.~{(\ref{ABCq})} and (\ref{ABC}) into
Eq.~{(\ref{Eq})}, we can obtain the numerical value for the
evolution functions, $E_q$s, at any given values of $Q^2$ and $x$. The results are shown in Fig.~{\ref{evo}}.

Using these functions we can evolve the sea quark densities to
higher values of energy scales. The results are depicted in
Fig.~{\ref{evosea}} at $Q^2=3\ \mathrm{GeV}^2$.

\section{Conclusions}
The polarized distribution function of mesons
cannot be supposed to vanish trivially, as otherwise  all sea
quark distributions in Eq.~{(\ref{seas})} would be zero. In addition,
there exist a variety of reliable studies that have calculated
the PPDF of mesons~\cite{lattice,ma,namkim}. We have determined these
polarized parton densities by calculating  the ratios of the
polarized valence densities inside the meson using Monte Carlo
algorithms and the expansion of the PPDFs in terms of quark masses (see
Eq.~(\ref{qmass})). In cases where the polarized valence
density of the pion is given (using any model), this method can
offer the corresponding functions for kaon and eta.

The orbital angular momentum was used to calculate the meson spins.
These equations are written and solved for the proton~\cite{ji}
to justify the spin crisis, and we solved them for mesons to
calculate the parton orbital angular momentum of mesons and justify
the zero spin of mesons. These calculations can be considered as additional
evidence for the existence of non-zero polarized valence distributions
for mesons. The agreement with the result in Ref.~\cite{MKY-2011}
confirms the validity of our calculations.

Due to the fact that the mass of the quarks can be responsible for
 chiral symmetry breaking, we employed the mass dependence of
the proton  quark densities, using the running mass equation
~\cite{ellis}, to reveal their asymmetry in a clearer way. The
functional form of the eigenvalues of the evolution operator could be
extracted, given appropriate boundary conditions for its
parameters. By numerical solution of the DGLAP evolution
equations, the numerical values of the required parameters in
Eq.(\ref{ABCq}) were obtained. At high enough energies, where the
quarks become massless, these eigenvalues tend to each other and
the degenerate formulation can be used.

For further research, the bare quark distributions inside the
proton can be obtained theoretically rather than
phenomenologically, based on the solution of the  Dirac equation
under a specified potential. The asymmetry of polarized light quark
distributions can also be investigated, considering the charge
asymmetry of parton densities, which we hope to work on in future.
\newpage
\section*{Appendix}

The coefficients in the function $a\, x^b (1-x)^c$  have a
typical expansion as  follows:
\begin{equation}
(a,\ b\ \mathrm{and}\ c)=\sum_{i=0}^3 R_i\ \alpha_s^i\;.
\label{para}
\end{equation}
Their numerical values for polarized valence, gluon and singlet
sector of pion, kaon and eta are calculated based on the valon
model. They  are
tabulated below. The $\alpha_s$ in Eq.~(\ref{para}) denotes the
running coupling constant at NLO approximation.  We have taken
$\Lambda_{\overline{MS}}=0.200\ GeV$ in all parts of these
calculations.
\begin{table}[h]
\begin{tabular}{|c|c|cccc|}
\cline{3-6}
\multicolumn{2}{c|}{} & $R_0$ & $R_1$ & $R_2$ & $R_3$ \\
\hline
                       & $a$ & 0.292 & 1.592 & -1.399 & 0.561 \\
\cline{2-2}
$\delta q_{v}^{\pi}$   & $b$ & 0.610 & 2.160 & -1.884 & 0.797 \\
\cline{2-2}
                       & $c$ & 2.628 & -4.435 & 4.940 & -2.215 \\
\hline
                       & $a$ & 0.229 & 3.784 & -2.899 & 0.983 \\
\cline{2-2}
$\delta \Sigma^{\pi}$  & $b$ & 0.0132 & 3.520 & -3.039 & 1.221 \\
\cline{2-2}
                       & $c$ & 2.404 & -3.509 & 3.664 & -1.632\\
\hline
                       & $a$ & -0.0540 & 4.832 & -7.796 & 5.515 \\
\cline{2-2}
$\delta g^{\pi}$       & $b$ & -0.598 & 3.220 & -4.905 & 4.505 \\
\cline{2-2}
                       & $c$ & 5.135 & -12.79 & 14.84 & -6.364 \\
\hline \hline
                       & $a$ & 0.269 & 1.564 & -1.302 & 0.522 \\
\cline{2-2}
$\delta q_{v}^K$       & $b$ & 0.586 & 2.098 & -1.840 & 0.778 \\
\cline{2-2}
                       & $c$ & 2.783 & -4.496 & 5.026 & -2.256 \\
\hline
                       & $a$ & 0.229 & 3.725 & -2.907 & 0.993 \\
\cline{2-2}
$\delta \Sigma^K$      & $b$ & 0.00802 & 3.455 & -3.019 & 1.212 \\
\cline{2-2}
                       & $c$ & 2.433 & -3.547 & 3.703 & -1.648 \\
\hline
                       & $a$ & -0.0423 & 4.688 & -7.443 & 5.212 \\
\cline{2-2}
$\delta g^K$           & $b$ & -0.592 & 3.129 & -4.692 & 4.298 \\
\cline{2-2}
                       & $c$ & 5.186 & -12.89 & 14.98 & -6.436 \\
\hline \hline
                       & $a$ & 0.245 & 1.527 & -1.193 & 0.478 \\
\cline{2-2}
$\delta q_{v}^{\eta}$  & $b$ & 0.553 & 2.008 & -1.777 & 0.751 \\
\cline{2-2}
                       & $c$ & 2.994 & -4.590 & 5.152 & -2.314 \\
\hline
                       & $a$ & 0.179 & 3.559 & -2.377 & 0.749 \\
\cline{2-2}
$\delta \Sigma^{\eta}$ & $b$ & -0.00478 & 3.358 & -2.988 & 1.195 \\
\cline{2-2}
                       & $c$ & 2.759 & -3.589 & 3.763 & -1.683 \\
\hline
                       & $a$ & -0.0954 & 4.478 & -7.264 & 5.417 \\
\cline{2-2}
$\delta g^{\eta}$      & $b$ & -0.582 & 2.933 & -4.177 & 3.836 \\
\cline{2-2}
                       & $c$ & 5.657 & -13.55 & 15.81 & -6.762 \\
\hline
\end{tabular}
\end{table}
\begin{multicols}{2}

\end{multicols}
\clearpage
\end{document}